\documentclass[useAMS,usenatbib]{mn2e}

\usepackage{psfrag}
\usepackage{color}
\usepackage{graphicx}

\title[Detecting a rotation in the $\epsilon$ Eridani debris disc]{Detecting a rotation in the $\epsilon$ Eridani debris disc}
\author[C. J. Poulton, J. S. Greaves and A. C. Cameron]{C. J. Poulton$^{1}$\thanks{E-mail: cp85@st-and.ac.uk}, J. S. Greaves$^{1}$ and A. C. Cameron$^{1}$\\\
$^{1}$School of Physics \& Astronomy, University of St. Andrews, North Haugh, St. Andrews, Fife KY16 9SS, UK}

\begin{document}

\date{Accepted 2005 . Received 2005 ; in original form 2005 }

\pagerange{\pageref{firstpage}--\pageref{lastpage}} \pubyear{2005}

\maketitle

\label{firstpage}

\begin{abstract}

The evidence for a rotation of the $\epsilon$ Eridani debris disc is
examined. Data at 850 $\micron$ wavelength were previously obtained using
the Submillimetre Common User Bolometer Array (SCUBA) over periods in
1997-1998 and 2000-2002.  By $\chi^2$ fitting after shift and rotation
operations, images from these two epochs were compared to recover proper
motion and orbital motion of the disc. The same procedures were then
performed on simulated images to estimate the accuracy of the results. 

Minima in the $\chi^2$ plots indicate a motion of the disc of
approximately 0.6$''$ per year in the direction of the star's proper motion.
This underestimates the true value of 1$''$ per year, implying that some
of the structure in the disc region is not associated with $\epsilon$
Eridani, originating instead from background galaxies. From the $\chi^2$
fitting for orbital motion, a counterclockwise rotation rate of
$\sim$2.75$^{\circ}$ per year is deduced. Comparisons with simulated data
in which the disc is not rotating show that noise and background galaxies
result in approximately Gaussian fluctuations with a standard deviation
$\pm$1.5$^{\circ}$ per year. Thus counterclockwise rotation of disc
features is supported at approximately a 2-$\sigma$ level, after a 4-year
time difference. This rate is faster than the Keplerian rate of 0.65$^{\circ}$
 per year for features at $\approx$65~AU from the star, suggesting
their motion is tracking a planet inside the dust ring. 

Future observations with SCUBA-2 can rule out no rotation of 
the $\epsilon$ Eridani dust clumps with $\sim$4$\sigma$ confidence. 
Assuming a rate of about 2.75$^{\circ}$ per year, the rotation of the 
features after a 10-year period could be shown to be $\geq$1$^{\circ}$ 
per year at the 3$\sigma$ level.

\end{abstract}


\section{Introduction}
\subsection{$\epsilon$ Eridani: A Special Case}

Debris discs around nearby stars represent extra-solar analogues of the
Kuiper Belt. It is in this context that studying the $\epsilon$ Eridani
dust ring is of particular interest since it is of spectral type K2V, not
too dissimilar to the Sun, but with a much younger age of 0.8 Gyr \citep{Song
2000, Di Folco 2004} and thus represents an analogue to the young solar
system. It is generally thought that protoplanetary discs evolve into
debris discs after planets have formed, a process that is completed
within 10-100 Myr after star birth \citep{Schutz 2004,Holland 1998}.
\citet{Greaves 2004} have noted that there is an apparent lack of
overlap between stars with debris discs detectable at submillimetre
wavelengths and those with radial velocity planet detections. There are
only a few stars with positive detections for both an infrared
dust-excess and a radial velocity planet \citep{Dominik 1998,Beichman
2005}. While this is likely to reflect the small number of stars searched
for both phenomena, $\epsilon$~Eridani remains unique in having resolved
structure within the debris ring \citep{Greaves 1998, Greaves 2005} plus
an inner gas giant planet detected by Doppler wobble \citep{Hatzes 2000}.
Further evidence for this planet comes from the forced offset of the
centre of the ring \citep{Greaves 2005}, consistent with a forced
eccentricity of the dust particles \citep{Wyatt 1999}. The ring structure
also appears perturbed as if by a more distant gas giant, but this body
has not yet been imaged, with a maximum mass constraint at around the
dust ring radius of approximately 5 M$_J$ \citep{Macintosh 2003}. 

\subsection{Planet Hunting by Tracking of Disc Features}

An infrared excess around $\epsilon$ Eridani was first detected during
photometric measurements by the Infrared Astronomical Satellite (IRAS)
\citep{Aumann 1988}, soon after the initial discovery of an IR excess
around Vega \citep{Aumann 1984}. Subsequent submillimetre observations at
850 $\micron$ with the Submillimetre Common User Bolometer Array (SCUBA)
by \citet{Greaves 1998} resolved the disc and showed a
ring-like structure. The faint ring surrounding $\epsilon$ Eridani
has not yet been successfully imaged at optical wavelengths 
\citep{Proffitt 2004}. 

Further SCUBA observations made in 2000-2002 \citep{Greaves 2005} allow
us to study the motions of the substructure in the ring over a 5 year
period. If clump features are tracking the orbital motion of a planet
within the ring, they should appear to rotate faster than the Keplerian
rate at the ring radius; this would be an unambiguous signature of a
planet with the forced period providing a measure of its orbital
semi-major axis.  Clumps will appear in a disc where planet formation has
taken place because the planets will migrate outwards via angular
momentum exchange with the remaining planetesimals. As the planets
migrate outwards, the associated gravitational resonances also move out,
thus trapping dust particles outside the planet's orbit into mean motion
resonances \citep{Wyatt 1999,Wyatt 2003}. The Kuiper Belt in the Solar
System probably formed in a similar way, where the material is thought to
have been pushed outwards by a mean motion resonance associated with
Neptune as it migrated further away from the Sun \citep{Gomes 2003,
Levison 2003,Malhotra 1995}. 

Circumstellar dust, trapped into resonances by an outwardly migrating
planet \citep{Wyatt 1999, Wyatt 2003, Quillen 2002}, is expected to be
observable at sub-millimetre wavelengths. Whilst the Vega and Fomalhaut
disc asymmetries have been modelled \citep{Wyatt 2003, Wyatt 2002}, the
$\epsilon$ Eridani dust ring offers a unique prospect to directly track
motion of clumps over time, because of its favourable inclination
($\sim$25$^{\circ}$ from face-on to the observer) and the fact that this
is one of the closest stars to the Sun at a distance of 3.22 pc.  Models
of the distribution of the $\epsilon$ Eridani clumps \citep{Quillen
2002,Ozernoy 2000} suggest the perturbing planet lies $\approx$40-60 AU
from the star, thus the clumps are expected to orbit with a period of
$\approx$280--520 years. These rotation rates of 0.7--1.3$^{\circ}$ per
year, corresponding to particular resonant periods \citep{Quillen
2002,Ozernoy 2000}, require long times to detect.  Here we present
a preliminary analysis over 5~years of data, making use of image fitting
to recover statistically any small rotation present. 

\section{Observations}

The results used here consist of the 5 year dataset presented by
\citet{Greaves 2005}. Observations of $\epsilon$ Eridani at 850 $\micron$
were made with SCUBA between 1997 August and 2002 December, in total
comprising 56 images and an integration time of 33.5 hours. The dataset
is split into two parts, one including data taken during 1997-1998 and
the other from 2000-2002; the observing runs were bunched in time so that
the effective mid-points of the two periods are around 4 years apart. The
2000-2002 dataset has a noise level lower by a factor of two, as a result
of longer duration and better sensitivity.  Data were also obtained at
450 $\micron$, but the SCUBA filter used in 1997-1998 had a low
throughput, and so this image can not be used for time-dependent studies. 

The analysis made here is of the co-added 1997-1998 data compared to the
co-added 2000-2002 results. The stellar proper motion is -1$''$ in RA per
year, so any disc emission associated with the star should shift by
approximately 4$''$ west, with some blurring because the observing
periods were spread out.  The diffraction-limited FWHM beam size at 850
$\micron$ was 15$''$ by 15.5$''$, but the data have been smoothed using a
7$''$ Gaussian to an effective 17$''$ beam \citep{Greaves 2005}.  The net
pointing errors are expected to be small, below the level of the $1''$
cell size in the images.  The noise for the 1997-1998 dataset is
4.3$\times$10$^{-3}$ mJy arcsec$^{-2}$ (0.96 mJy beam$^{-1}$) with
2.4$\times$10$^{-3}$ mJy arcsec$^{-2}$ (0.54 mJy beam$^{-1}$) for the
2000-2002 dataset. Photospheric emission of 1.7$\pm$0.2 mJy has been
subtracted from the images, for ease of comparison with simulations that
neglect the star.

\section{Simulated Data}

The proper motion and rotation of dust features associated with the star
can be assessed by translating and rotating the 1997-1998 and 2000-2002 co-added 
images to find the best match.  The accuracy of the results was evaluated
by comparison with simulated data with comparable noise, background
galaxies and a foreground consisting of clumps embedded in a smooth ring
of radius 20$''$ centred on the star.

Each simulated image was constructed from a set of frames, each
representing an individual observation taken with SCUBA. The simulated
analogue of the 1997-1998 dataset comprised 22 frames of equal depth,
which when co-added reproduced the noise in the actual data at the final
17$''$ resolution. The range of proper motion offset compared to the last
real data taken was 4-5$''$. A similar procedure was followed to simulate
the 2000-2002 data, using 34 frames and a proper motion range of
0--2.5$''$. The difference in depths between the real 1997-1998 and
2000-2002 datasets was accounted for by both the number of images in
 each simulated dataset and the noise levels that were
input into each set. While the number of input frames is the real value,
the equal noise value per frame is a simplification (as it was too
complex to simulate the duration and observing conditions of every real
frame). Thus the effective mid-points of the two observing periods, about
4 years apart, may not be exactly matched in the simulated results. 

Each simulated frame representing an observation by SCUBA was
constructed from random noise, random background galaxies and a foreground
(ring and clumps embedded in the ring) of comparable brightness to the
observed data.  The same background galaxies were used in both the
1997-1998 and 2000-2002 frames but a different sample generated for each
simulation. The chopping motion of the secondary mirror of the telescope
used to determine sky levels was also simulated. All of these effects
contribute to the real images; the data should thus represent one
possible outcome of the simulations, to within the accuracy of the
simplifications used. 

\subsection{Simulation details}

The noise image was made by choosing the flux for each individual pixel
randomly from a Gaussian distribution and then smoothing spatially with a
7$''$ Gaussian as was done for the observed data. This assumes the noise
per 1$''$ pixel is statistically independent.  The
mean for the noise distribution was zero and the $\sigma$ value was
chosen so that the final simulated images had a standard deviation of
flux per beam matching the real dataset once the galaxy population had been added. 

The integral galaxy counts N($>$S) for each flux, S, at 850 $\micron$
were modelled with Poisson statistics using numbers from \citet{Barnard
2004}, by a double power law given by

\begin{equation}
N(>S) \propto S^{-\alpha}
\end{equation}

\noindent with $\alpha$=0.94 for S $<$ 1.18 mJy and $\alpha$=2 for S $\geq$ 1.18
mJy. Each galaxy's point-like flux was smoothed with a 2-D Gaussian to reflect
the size of the beam of the JCMT at 850 $\micron$ and then placed at
random coordinates chosen from a uniform distribution. Thus no area of
the image was favoured and any possible galaxy clustering was neglected.

The foreground was modelled by adding beam-sized 2-D Gaussian regions of
flux to a ring.  The ring was created by placing Gaussians at every pixel
at distances between 17.5$''$-22.5$''$ from the star, producing an
annulus centred on the star with the radius and width observed
\citep{Greaves 2005}.  The flux per pixel in the ring was chosen to match
the smooth level observed (i.e. between clumps) of $\approx$10$^{-2}$ mJy
arcsec$^{-2}$. \citet{Greaves 2005} identified three clumps with possible
rotation located northeast, northwest and southeast of the star at a
radius of 20$''$ and thus three clumps were added at these positions in
the simulated ring. The peak fluxes of the clumps were set at
$\approx$1.5 x 10$^{-2}$ mJy arcsec$^{-2}$ so that the total foreground
had peak fluxes at the clump locations of $\approx$2.5 x 10$^{-2}$ mJy
arcsec$^{-2}$. This is the mean of the total fluxes towards the three
candidate moving clumps in the observed data \citep{Greaves 2005}. 

Proper motion of the disc was simulated by translating the foreground
relative to the background for each of the simulated observations, by a
distance corresponding to the epochs of individual real frames.  In the
observed data, a correction for the annual proper motion of $\epsilon$
Eridani ($\mu_\alpha$ = -0.976$''$ yr$^{-1}$, $\mu_\delta$ = 0.018$''$ yr$^{-1}$)
\citep{ESA 1997} was made by sorting and
shifting the frames to a precision of 0.5$''$ bins in R.A. only \citep{Greaves 2005}. 
This correction was also performed on the simulated observations, and so
the background galaxies (fixed with respect to the sky) will appear to
move at the rate of the proper motion but towards positive R.A., in a
co-ordinate frame co-moving with the star. 

Also simulated was the chopping procedure performed by the JCMT, used to
remove sky fluctuations. `Blank' regions of sky on either side of the
field of interest are observed interleaved with the on-source
observations, and subtracted to leave only astronomical signal; for
comments on the limitations see \citet{Archibald 2002}. If background
galaxies lie within the SCUBA field of view at either off-source
position, ``holes'' with magnitude equal to half the brightness of the
galaxies being chopped on to will appear in the final observed image. Due
to variations in the chop direction, such holes will be blurred in the
on-source residual frame and thus difficult to recognise. As chopping
introduces extra fluctuations and modifies the background galaxy
contributions, it was important to include it in the simulations. 

The noise, background galaxies and foreground images were constructed for a 400
by 400 pixel square frame but the field of view of SCUBA on the JCMT is
roughly circular with a radius of 72$''$.  It was necessary to simulate
this larger area of sky to allow the chopping procedure to be simulated. 
Flux values for each pixel within an area of sky were read into 2-D arrays
at the position of the source, M, and the two off-source positions to
the left, L, and right, R, of the source.  The pixel fluxes used in the
final images, I, were then calculated using

\begin{equation}
I(x,y) = M(x,y) - {(L(x,y) + R(x,y)) \over 2}.
\end{equation}

\noindent Chopping in azimuth was used for the real data, so that the
left and right chop throws fell on different sky regions depending on the
source elevation. This was accounted for in the simulations by varying
the angle between the chop throw and the line of constant right ascension
(horizontal in the images) randomly between -50$^{\circ}$ and
50$^{\circ}$. In the observations, chop throws of 80$''$(5\% of images),
 100$''$(55\%), and 120$''$(40\%) were used. A chop throw of 120$''$ was
 used for all of the simulated chopping. 

Some small effects were neglected in the simulations. The disc is fairly close
 to face-on to the observer, but inclined at $\sim$25$^{\circ}$
 from the sky plane and so actually appears as an ellipse in the sky.
 Assuming a rotation rate of 2.75$^{\circ}$ per year, a clump moving along
 the direction of the minor axis of the ellipse will have a motion that
 is only $\sim$0.3$''$ less than a clump moving parallel to the direction of the
 major axis. Therefore, it was reasonable to approximate the ring as 
circular to simplify the rotation of the foreground image. The annual shifts in position
 due to parallax for $\epsilon$ Eridani are $\pm$0.3$''$ and these were neither
 accounted for in the telescope tracking software nor included in the
 modelling. In the observed data 68\% of the images co-added were taken
 between August-October and so the effect of smearing due to parallax errors
 will be less than 0.3$''$. The dust ring was also simulated as being centred on the star,
 neglecting the offset of 1.5-2$''$ identified by \citet{Greaves 2005} and
 suspected to be forced by the inner planet. We assume the bulk motions of
 the clumps still follow circular orbits. 

\begin{figure*}
\centerline{
\includegraphics[width=3.0in]{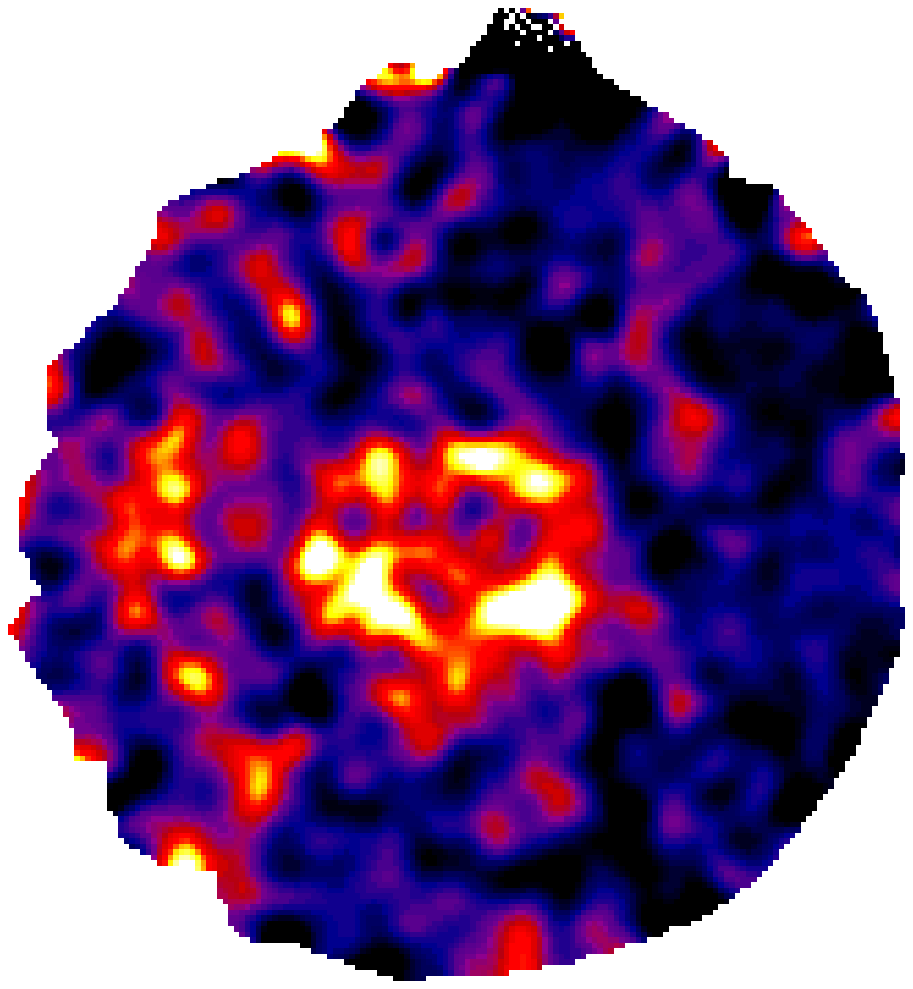}
\includegraphics[width=3.0in]{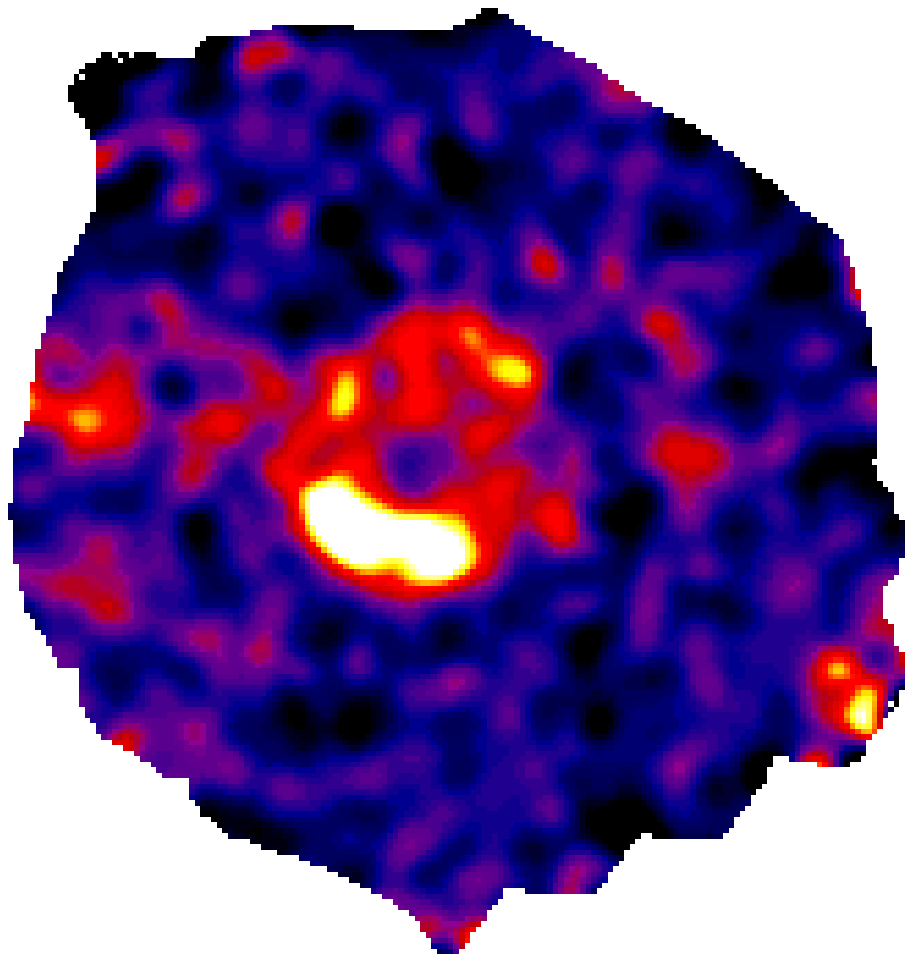}
}
\normalsize
\caption[]{Observations of the $\epsilon$ Eridani debris disc taken with
SCUBA on the JCMT. (Left) 1997-1998 dataset, (averaged over 22 images) in 
a 1$''$ pixel grid in R.A., Dec coordinates (north is up, east is
left), plotted with the dynamic ranges min=-1$\times$10$^{-2}$,
max=2.5$\times$10$^{-2}$ mJy arcsec$^{-2}$. (Right) 2000-2002 dataset 
(averaged over 34 images) with the same parameters.}
\end{figure*}

\begin{figure*}
\centerline{
\includegraphics[width=3.0in]{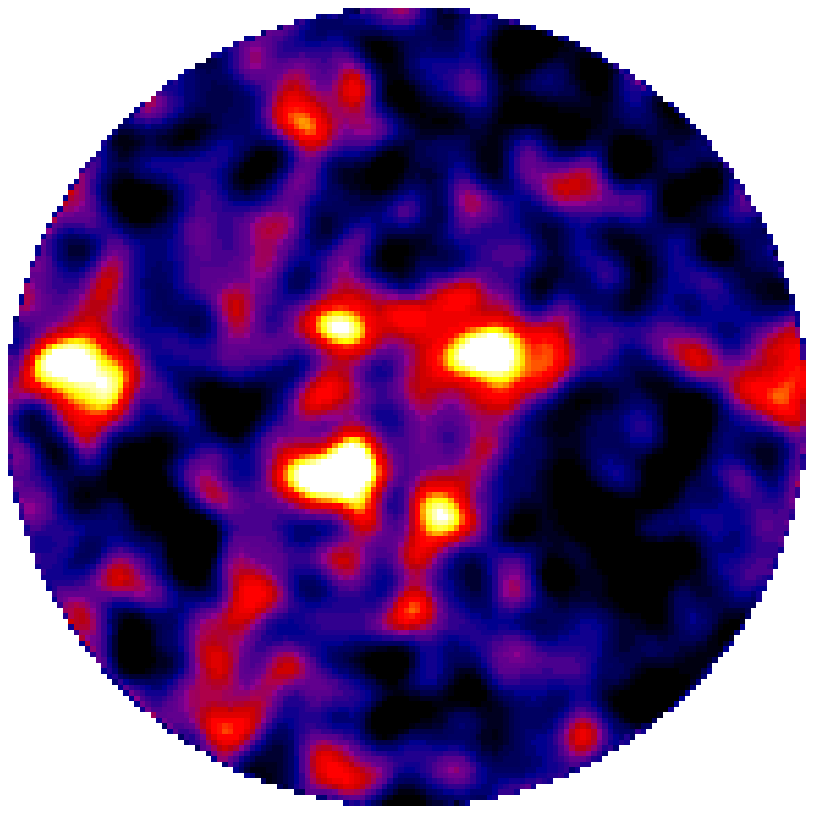}
\includegraphics[width=3.0in]{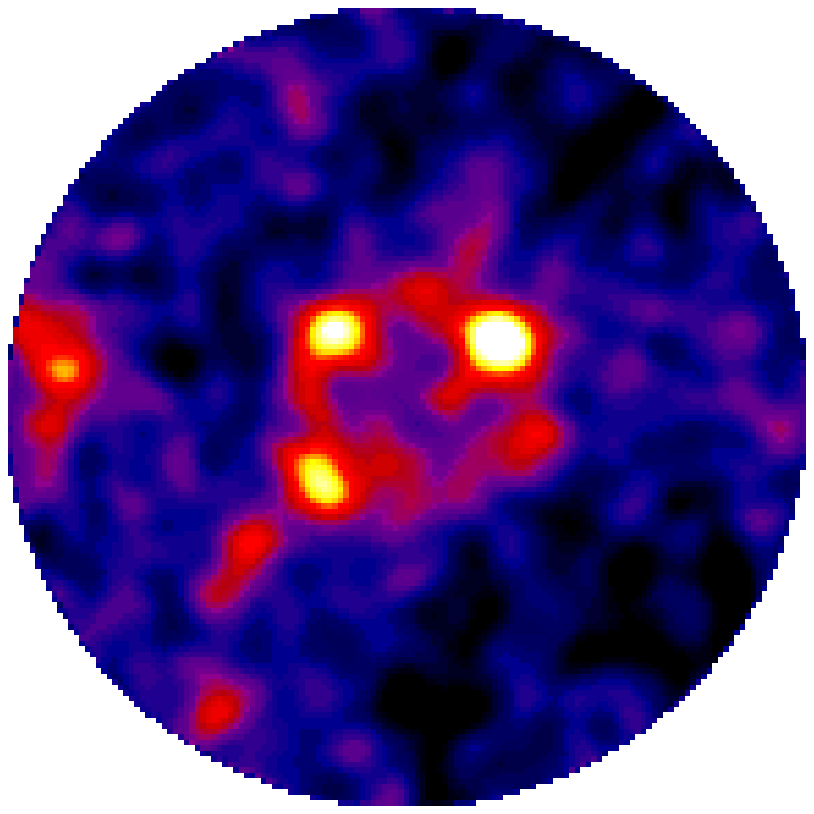}
}
\normalsize
\caption[]{As for Figure 1 but for simulated  SCUBA images with no 
rotation of the foreground clumps: (left) example of simulated 1997-1998 
image, (right) the corresponding 2000-2002 image. Other parameters are as 
in Figure~1.}
\end{figure*}

\subsection{Simulation output}

The real images are shown in Figure~1 and examples of the simulated
images are shown in Figure 2. The simulated images recreate a
ring with comparable brightness, size and morphology to that seen in the
observations. The regions surrounding the ring also show comparable
levels of noise and density of background objects. Differences between the two
simulated images are at a similar level to those in the real data of
Figure~1. Although the image structure looks complex, this is in part
because the false-colour scale both includes a wide flux range (negative
as well as positive signals) and also tends to enhance features with low
levels of contrast, as discussed by \citet{Greaves 2005}. 

The simulation procedure was also repeated to resemble a future
observation that could be taken in 2007 with SCUBA-2 -- a
replacement camera for SCUBA with greater sensitivity, fidelity and field
of view \citep{Audley 2004, Holland 2003}. For this simulation, the noise was set to
zero as it will fall below the background confusion in an observation of
only $\sim$1 hour.  The chopping part of the simulated data algorithms
was also disabled since SCUBA-2 will sample rapidly enough that sky
fluctuations will be negligible and so chopping will not be required.
(Chopping can be re-introduced in the data analysis where necessary for
matching SCUBA-2 images of $\epsilon$ Eridani with those already taken
with SCUBA.) Any candidate rotation of the disc identified here can thus
be checked by extending the total timeline of observations to 10 years
(from 1997 to 2007). 

\section{Detecting a Rotation}

\begin{figure*}
\centerline{
\includegraphics[width=3.5in]{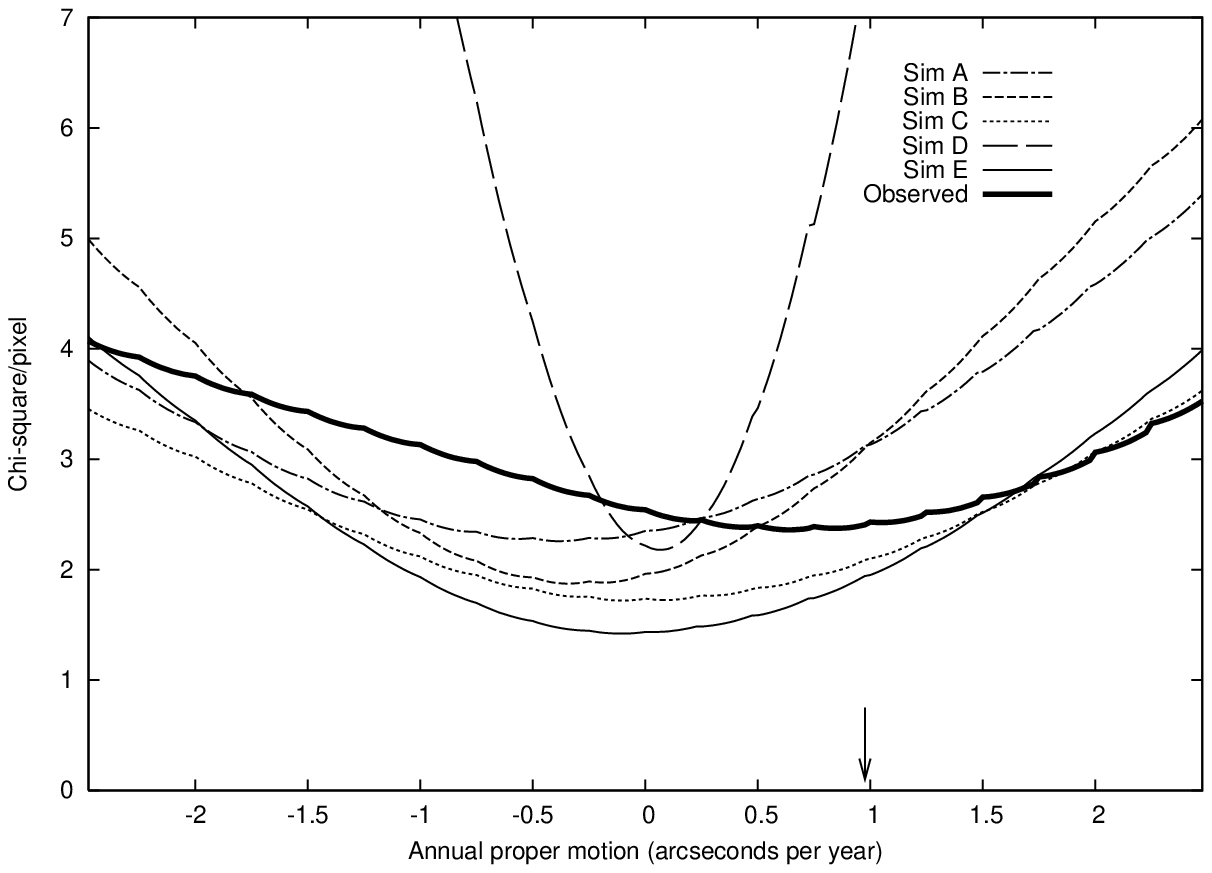}
\includegraphics[width=3.5in]{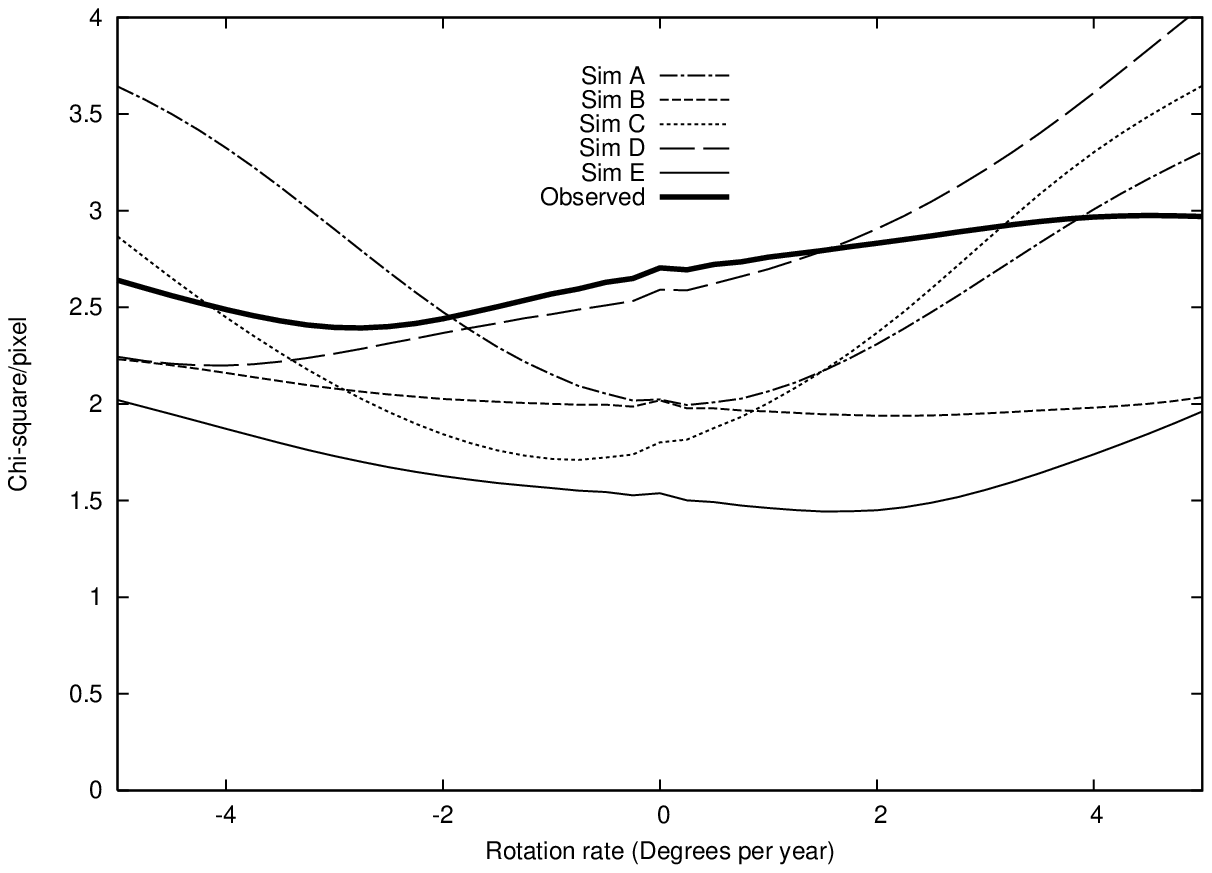}
}
\normalsize
\caption[]{$\chi^2$ per pixel curves for observed data and simulations.
(Left) Fits for annual proper motion shift from the 1997-1998 to the 
2000-2002 dataset. Thick curve shows observed data; other curves are 5
random simulations with no proper motion or rotation. Positive shifts
are to the right (-ve in R.A.) and the arrow indicates the known proper motion 
of $\epsilon$ Eridani ($\mu_\alpha$ = -0.976$''$ yr$^{-1}$). Small deviations in
the curves are caused by slight imperfections in the linear 
interpolation. The narrow curve (Sim D) originates from a very bright
galaxy in the ring, forcing a minimum that must be close to zero. (Right)
Fits for annual rotation rates, comparing 1997-1998 and 2000-2002
datasets and with corrections made for proper motion. Thick curve shows
observed data; thin curves are 5 random simulation pairs with proper
motion corrections included but no rotation.}
\end{figure*}

The simplest method of measuring rotation would be to track individually
identified clumps. However, clump extraction algorithms \citep{Williams
1994,Bertin 1996} are not well adapted to finding clumps and arcs within
a ring. Hence a $\chi^{2}$-fitting technique was adopted instead, with
the advantage of using all the information in the image simultaneously.
The $\chi^{2}$ per pixel value for two images, I$_1$ and I$_2$, with
measured noise $\sigma_1$ and $\sigma_2$, using only pixels in a box with
its lower left and upper right corners defined by (x,y) cartesian
co-ordinates $(a_1,b_1)$ and $(a_2,b_2)$, can be calculated using

\begin{equation}
\chi^2 = {\sum_{x=a_1}^{a_2}\sum_{y=b_1}^{b_2}{(I_1(x,y) - I_2(x,y))^2} \over (a_2-a_1+1)(b_2-b_1+1)(\sigma_1^2+\sigma_2^2)}.
\end{equation}

\noindent The $\chi^{2}$ per pixel values between the 2000-2002 dataset
and the 1997-1998 dataset were calculated as the 1997-1998 dataset was
shifted in right ascension and rotated relative to the 2000-2002 dataset,
using the tasks ``SLIDE'' and ``ROTATE'' in the KAPPA software package
\citep{Currie 2004}. The minima in plots of $\chi^2$ per pixel as
a function of right ascension shift or rotation angle then identified the
best solutions for each (independently).  The region of the image
included in the $\chi^2$ fit was defined by a 70$''$-square box, as this
included all of the disc with as little off-source background as
possible; extra background galaxies included in a larger box tend to
cause the $\chi^2$ fits to incorrectly favour a shift and rotation of
zero.

For the simulated SCUBA-2 observation, the $\chi^2$ per pixel values were
calulated twice, once for the future 2007 observation compared to the
1997-1998 SCUBA dataset, and again for 2007 compared to the 2000-2002
SCUBA data. The two $\chi^2$ values corresponding to the same shift or
rotation rate were then averaged to create the final $\chi^2$ per pixel
curves. We also performed simulations with no galaxies or
noise, and verified that the true rotation values were recovered
correctly. 

There are three possible ways in which the regions of emission in the
ring may be behaving:
\begin{enumerate}
\item The ring consists of only background galaxies and therefore the
features are not tracking with the proper motion of the star or revolving
around the star's position.
\item The ring features are tracking with the proper motion of the star
but are revolving around the star at the Keplerian period of the ring
 which would not require a planet to explain the motion.
\item The ring features are tracking with the proper motion of the star
and revolving around the star at the period of a planet at tens of AU.
\end{enumerate}

To establish which of these explanations gives the best description of
the data, the observed data were fitted for the proper motion of 
emission in the disc region, and then fitted to establish a rotation 
rate of the ring features. Then for comparison, 100 sets of
simulated data were made with no proper motion or rotation of the
foreground, to quantify the random fluctuations due to noise, background
galaxies and chopping on to background galaxies. This was then repeated
but assuming the foreground was tracking with the proper motion of the
star and with the foreground rotated at rates of 0$^{\circ}$ (i.e. no
rotation), 1$^{\circ}$, 2$^{\circ}$, 4$^{\circ}$ and 10$^{\circ}$ per
year (unrealistically large but included for comparison). The purpose of
trying different rotation rates was to examine how well the true rate is
recovered, given that non-moving background galaxies introduce a bias
towards a $\chi^2$ minimum around zero degrees. 

\section{Results and Discussion}

\begin{table*}
\centering
\begin{minipage}{140mm}
  \caption{Proper motion shifts and rotations from $\chi^2$ fitting of 6 simulated SCUBA datasets each consisting of 100 simulated 1997-1998 and 2000-2002 images with foreground background galaxies, noise and chopping included.}
  \begin{tabular}{@{}ccccccccc@{}}\\
  \hline
  \hline
Simulation & Simulated & Average & Motion & Simulated & Average & Rotation & 5$\sigma$ Clipping & Average \\
Set No. & Motion\footnote{Positive values signify a direction negative in R.A.} & Motion & $\sigma$ & Rotation & Rotation & $\sigma$ & Results & Min $\chi^2$ Value \\
& ($''$ yr$^{-1}$)  & ($''$ yr$^{-1}$) & ($''$ yr$^{-1}$) &  (${^\circ}$ yr$^{-1}$) & ($^{\circ}$ yr$^{-1}$) & ($^{\circ}$ yr$^{-1}$) & (No. Excluded) \\
\hline
1 & 0 & 0.01 & $\pm$0.28 & 0 & -0.03 & $\pm$1.23 & 5 & 1.92 \\
2 & +1 & 0.79 & $\pm$0.38 & 0 & -0.15 & $\pm$1.58 & 4 & 2.03 \\
3 & +1 & 0.78 & $\pm$0.42 & -1 & -0.68 & $\pm$1.83 & 5 & 2.03 \\
4 & +1 & 0.73 & $\pm$0.44 & -2 & -0.96 & $\pm$1.87 & 2 & 2.06 \\
5 & +1 & 0.65 & $\pm$0.47 & -4 & -1.92 & $\pm$2.55 & 7\footnote{4$\sigma$ clipping used for these values} & 2.14 \\
6 & +1 & 0.63 & $\pm$0.47 & -10 & -4.83 & $\pm$13.71 & -\footnote{$\sigma$ clipping not possible due to broad distribution of results} & 2.42 \\
\hline
\end{tabular}
\end{minipage}
\end{table*}

Figure 3 shows $\chi^2$ per pixel as a function of the annual proper motion 
and rotation rate of the 1997-1998 image relative to the 2000-2002 data. 5 examples of 
simulations are shown for comparison. Table 1 lists the statistics from the full  
sets of simulations, each set comprising 100 comparisons of pairs of 
images. 

\subsection{Motion of the Disc}

Figure~3 (left panel) shows that in the observed data, the emission in
the ring region has shifted at a rate of 0.6$''$ per year to the right
(negative in R.A.) when tracked from the 1997-1998 to the 2000-2002
epoch.  \citet{Greaves 2005} argued that at least some of the disc
features are tracking with the proper motion of the star, i.e. 1.0$''$/yr
to the right. The fact that the net motion measured is smaller may be
explained by the presence of stationary background features behind the
ring, which by themselves would produce a minimum in the $\chi^2$ curve
at $0''$ per year. A result intermediate between 0$''$/yr and 1$''$/yr
thus indicates a mixture of tracking and stationary features. The net
value of $\approx$0.6$''$/yr suggests that the majority of features in
the ring region are truly associated with $\epsilon$ Eridani.

The simulations (Table~1) reproduce this result closely. An input proper
motion of 1$''$/yr is recovered as on average $\approx$0.7$''$/yr of net
motion (with little dependence on any assumed ring rotation). Thus the
stationary galaxies do in fact reduce the net motion measured in the same
sense as inferred in the real data. The standard deviation of the proper
motion solutions is approximately $\pm$0.4$''$/yr. Thus the real-data
solution of 0.6$''$/yr, differing by only $\approx$0.1$''$/yr from the 
typical simulation result, is well within the scatter. 

Further tests confirmed the validity of the simulations. When the
observed data were corrected for the proper motion of the star, the
solution obtained was 0.4$''$ per year to the left (positive in R.A.).
This is a measure of the amount of emission that is fixed in the sky and
so moving to the left in the co-ordinate frame of the star. This is the
value expected, as the difference between the 1$''$/yr if all the
emission moved with the star and the 0.6$''$ actually measured. 
A set of simulations was also performed to check that
a non-moving disc returned the correct result of 0$''$ per year. The
measured mean was very close to zero (Table~1) with a standard deviation
of $\pm$0.25$''$/yr. This is a measure of the fluctuations that can occur
due to random noise and chopping onto background galaxies; it is somewhat
smaller than the $\approx$0.4$''$ obtained in the more complex cases with
both stationary and moving flux contributions. Finally, we verified that
the small `ringing' effect seen in the curves in Figure~3 does not
affect the position of the minimum in the $\chi^2$ curves. 
This ringing arises from a
small jump at integer values, apparently in the bilinear interpolation of
the ``SLIDE'' task in KAPPA. Plotting the curves using only integer
values changed the average motions recovered by only $\sim$0.01$''$.

\subsection{Rotation of the Disc}

\begin{table*}
\centering
\begin{minipage}{140mm}
  \caption{Proper motion shifts and rotations from $\chi^2$ fitting of 6 
   simulated datasets including the SCUBA frames of Table~1 and 
   a SCUBA-2 image to be obtained in 2007 (for which no noise
   is included).}   
  \begin{tabular}{@{}ccccccccc@{}}\\
  \hline
  \hline
Simulation & Simulated & Average & Motion & Simulated & Average & Rotation & 5$\sigma$ Clipping & Average \\
Set No. & Motion\footnote{Positive values signify a direction negative in R.A.} & Motion & $\sigma$ & Rotation & Rotation & $\sigma$ & Results &  Min $\chi^2$ Value \\
& ($''$ yr$^{-1}$)  & ($''$ yr$^{-1}$) & ($''$ yr$^{-1}$) &  (${^\circ}$ yr$^{-1}$) & ($^{\circ}$ yr$^{-1}$) & ($^{\circ}$ yr$^{-1}$) & (No. Excluded) \\
\hline
1 & 0 & 0.00 & $\pm$0.07 & 0 & -0.05 & $\pm$0.26 & 0 & 2.05 \\
2 & +1 & 0.81 & $\pm$0.21 & 0 & 0.05 & $\pm$0.64 & 0 & 3.07 \\
3 & +1 & 0.78 & $\pm$0.22 & -1 & -0.52 & $\pm$0.68 & 0 & 3.19 \\
4 & +1 & 0.75 & $\pm$0.22 & -2 & -1.04 & $\pm$1.08 & 1 & 3.11 \\
5 & +1 & 0.76 & $\pm$0.25 & -4 & -2.48 & $\pm$1.79 & 1 & 3.44 \\
6 & +1 & 0.76 & $\pm$0.25 & -10 & -3.70 & $\pm$6.00 & -\footnote{$\sigma$ clipping not possible due to broad distribution of results} & 4.52 \\
\hline
\end{tabular}
\end{minipage}
\end{table*}

The $\chi^2$ fit between the observed 1997-1998 and 2000-2002 datasets
(corrected for proper motion) shows a rotation of $\approx$2.75$^{\circ}$
per year counterclockwise (see Figure 3, right panel). This is in the
same direction as but somewhat larger than the rotation of
$\sim$1.5$^{\circ}$ suggested by \citet{Greaves 2005}. It is also larger
than the rotation rate of 0.7$^{\circ}$-1.3$^{\circ}$ per year predicted
by \citet{Ozernoy 2000} and \citet{Quillen 2002}, assuming dust is trapped in
resonances with a planet at $\approx$40-60 AU from the star.

The simulated data are used as a guide to what rotation rates can be
confirmed or ruled out. In the simulation results (Table~1), the $\chi^2$
fit results from sets of 100 image-pairs showed that the rotations
obtained were approximately Gaussian distributed, for the cases where the
rotation rate was 0$^{\circ}$, 1$^{\circ}$ and 2$^{\circ}$ per year.
There were, however, a few spurious results which involved an unphysical
disc feature rotation rate (total rotations of $>$50$^{\circ}$). These
results occurred when a bright clump was randomly generated by a
combination of noise and background galaxies and fortuitously positioned
in the dust ring to give a deeper minimum in the $\chi^2$ curve
corresponding to a large rotation angle. These outcomes ($\leq$ 5 per set
of 100 simulations) did not fit into the Gaussian distribution and were
removed by clipping at the 5$\sigma$ level. After removing these points,
the sample standard deviation was typically $\sim$1.5$^{\circ}$ of
rotation per year. For rotation rates of 4$^{\circ}$ and 10$^{\circ}$ per
year, the rotations obtained followed a broader
distribution, and so no reliable measure of the rotation was obtained.

The true rotation rate is generally not recovered in the simulations,
with an average annual rotation measured generally about half of the
input rotation. This can be understood as the effect of the stationary
galaxies, which will introduce a bias towards a solution of zero.
 (When no rotation is input, a solution close to zero is in fact
recovered.) Only simulations in roughly the -1$\sigma$ tails of the
distributions recover the true input rotation rate (Table~1). There is
therefore a moderate probability that the rate measured in  the real
data, of $\approx$2.75$^{\circ}$ per year, is actually an underestimate
for the true value.

Table 1 shows that for the simulated data with no rotation of the
foreground, the sample standard deviation was 1.44$^{\circ}$ per year. 
As the measured rotation was 2.75$^{\circ}$ per year, this means that the
null hypothesis (zero or very slow ring rotation) is ruled out at the
$\approx$2$\sigma$ level. However, the actual rotation rate can not be
extracted from comparison with the simulations as the results overlap;
that is, the standard deviations are wider than the changes in mean value
between sets with different input rotation rates. Some estimate can be 
made of how plausible a measured rotation rate of 2.75$^{\circ}$ 
anticlockwise is within each simulation set. We find that the probability
of measuring at least this large a counterclockwise rotation is 10-25\% 
for rotation rates of 1-4$^{\circ}$ per year.

Tests were made to see if any further observational constraints reduced 
the simulation parameter space. In the real data, the brightest signal is 
10 mJy within one beam. Galaxies brighter than this within the 70$''$ 
box are predicted to occur in only $\approx$5\% of the simulations. 
Eliminating these might reduce the tendency towards solutions of zero 
rotation, but in fact an increase of approximately 20\% was the largest
change to any recovered rotation rate when this was done.

We also verified that small defects seen in the curves about zero
 probably arising from interpolation in the rotation process
did not significantly change the results. Removal of the defects
only changed the rotation rate recovered of the non moving non rotating case and only by 
$\sim$0.001$^{\circ}$ per year and the associated standard deviation
 by only $\sim$0.004$^{\circ}$ per year.

\subsection{Future Observations}

Table 2 shows the results for the simulated datasets with a third image
with no noise included.  This is intended to reflect observations that
 could be made with SCUBA-2 at 850$\micron$ in 2007, with noise that is
 negligible compared to random background galaxies. Chopping was
 re-introduced to allow matching of SCUBA and SCUBA-2 images, although
testing showed that this only sigificantly reduced the random fluctuations
in the minima in the 4$^{\circ}$ and 10$^{\circ}$ per year cases. The
 $\chi^2$ minima were found by comparing this simulated SCUBA-2 dataset
 with the two simulated SCUBA predecessors, as described above.
 The projected 10-year timeline should refine the solutions for the
 disc motions, and so the annual errors are expected to be reduced.

The proper motion is now recovered at similar mean values to before
($\approx$0.7$''$ per year) but with lower standard deviations. The
errors of $\approx \pm$0.25$''$/yr are similar to the SCUBA-only
non-moving simulation, suggesting this is a limit set by the fluctuations
within the two SCUBA datasets from 1997-1998 and 2000-2002. The input
rotation is not recovered more reliably than before but the dispersion
of the results is smaller, significantly so in the cases of up to
2$^{\circ}$/yr rotation. The number of doubtful results with very large
rotations is also reduced (Table~2). However, the distributions
of rotation values still overlap between different input rates, and so
addition of a third-epoch SCUBA-2 image would not uniquely identify the
true rotation rate. In fact, eliminating noise by comparing two
hypothetical future SCUBA-2 observations, with mid-points 4 years apart,
still did not resolve this ambiguity. We conclude that $\chi^2$ fitting
alone can not identify an orbital period; the paradox inherent in trying
to fit both moving and non-moving components with one rotation rate is
likely to be responsible. 

With the inclusion of the SCUBA-2 epoch, the case of no rotation could be
ruled out at the $\approx$4$\sigma$ level.  The average rotation rate
recovered was +0.04$^{\circ}$ per year with a sample standard deviation
$\pm$0.74$^{\circ}$. Hence if the rotation rate of $-2.75^{\circ}$ per
year estimated from the present data were to persist, this would be
around the -4$\sigma$ bound of this simulation.  Further, a rotation rate
of $\geq$1$^{\circ}$/yr would be confirmed at around the 3$\sigma$ level:
that is, a continuing measurement of $-2.75^{\circ}$ per year would be just
within the -3$\sigma$ tail of the simulation with an input rate of
1$^{\circ}$/yr. This would support clumps orbiting faster than the
Keplerian rate at $\approx$65~AU radius, which is 0.65$^{\circ}$ per year
for a 0.9~M$_{\odot}$ star. 

It may also be possible to observe at 450 $\micron$ with SCUBA-2 and the
improved resolution of 8$''$ at this wavelength may allow the rotation to
be constrained more accurately. This would require two epochs of SCUBA-2
data a few years apart, and also a better knowledge of the galaxy
counts at 450 $\micron$.

\section{Conclusions}

We have identified a motion of the $\epsilon$ Eridani disc consistent
with features tracking in the same direction as the proper motion of the
star. The best fitting proper motion $\mu$ = 0.6$''\pm$0.4$''$ per year is most
probably less than the star's proper motion of 1$''$ per year, confirming
that some of the features in the vicinity of the disc are likely to be
background galaxies. The measured rotation rate of 2.75$^{\circ}$ per
year counterclockwise is in the same direction but nearly twice as large
as that suggested by \citet{Greaves 2005}; it could also be an
underestimate as stationary galaxies tend to bias the rotation solution
below the real amplitude. Comparisons with simulated data show that the
measured rotation is significant at the $\sim$2$\sigma$ level, and a
future image with SCUBA-2 in 2007 could rule out no or very slow rotation
at the $\sim$4$\sigma$ level. The technique of $\chi^2$ fitting is
inherently limited by trying to find one solution that fits both
stationary galaxies and moving ring features, hence a method that 
identifies individual clumps or a method capable of using the proper motion 
to distinguish the foreground and background images will be needed to 
measure the true rotation rate. 

Radial velocity detections of a planet are restricted to finding planets
out to only a few AU from the star. In the case of $\epsilon$ Eridani,
finding a planet out to tens of AU via radial velocity measurements will
be extremely difficult since the period of the planet is deduced to be
$>$100 years, and the magnitude of the Doppler wobble is reduced by the
nearly face-on inclination ($\sim$25$^{\circ}$) of the disc. However,
this viewing angle also means that $\epsilon$ Eridani provides a unique
opportunity to track the motion of the substructure within the disc. 
This is the first ever analysis attempting to track clumps rotating in a
dusty disc. Future observations with SCUBA-2 can confirm the rotation of
clumps in resonance with a planet at tens of AU, while imaging with
submillimeter interferometers such as SMA and ALMA could radically
shorten the timescale to identify such motions.

\section*{Acknowledgments} 

CJP would like to thank Paul Clark and Clare Dobbs for helping with
 Latex questions. The JCMT is operated by the Joint Astronomy Centre,
 on behalf of the UK Particle Physics and Astronomy Research Council,
 the Netherlands Organisation for Pure Research, and the National
 Research Council of Canada.

\bsp

\label{lastpage}

\end{document}